\documentclass[aps,english,prl,twocolumn]{revtex4}
\usepackage{mathrsfs}
\usepackage{graphicx}
\usepackage{amsmath, amssymb}
\usepackage{babel}

\begin{document}

 \title{Ginzburg-Landau theory of dirty two band $s_{\pm}$ superconductors}
 \author{Tai-Kai Ng}

 \begin{abstract}
  In this paper we study the effect of non-magnetic impurities on two-band $s_{\pm}$ superconductors by deriving the corresponding
  Ginzburg-Landau (GL) equation. Depending on the strength of (impurity-induced) inter-band
  scattering we find that there are two distinctive regions where the superconductors behave very differently. In the
  strong impurity induced inter-band scattering regime $T_c<<\tau^{-1}_t$, where $\tau_t\sim$ mean-life time an electron stays in one
  band the two-band superconductor behaves as an effective one-band dirty superconductor. In the other limit $T_c\geq\tau^{-1}_t$,
  the dirty two-band superconductor is described by a network of frustrated two-band superconductor grains connected by
  Josepshon tunnelling junctions. We argue that most pnictide superconductors are in the later regime.
 \end{abstract}

\maketitle

  With the discovery of the Iron-based (pnictides) superconductors, superconductivity characterized by more than one
 order parameters, i.e. the multi-gap superconductors, becomes a topic of interests. Band structure calculations indicate
 that the materials have a quasi-two-dimensional electronic structure, with five bands centered around the $\Gamma$- and
 $M$- points in the Brillouin zone contributing to the Fermi surface.  It has been proposed that the superconducting
 order parameters in this multi-band materials has so called $s_{\pm}$-wave symmetry, where the order parameters have
 $s$-symmetry but with opposite sign between bands centered at $\Gamma$- and $M$-points\cite{hu,mazin,wang}.

  The effect of impurities in this class of materials has been an issue of interests. NMR\cite{NMR1} and
  lower critical field data\cite{field} seems to suggest the existence of nodes in the superconducting order parameter while
  APRES experiemnts\cite{ARPES1,ARPES2} favor node-less gaps. One possible solution to this
  controversy is that large number of in gap states are induced by impurities in the material because of the special
  $s_{\pm}$ order-parameters. Indeed, such a scenario has received supports from self-consistent-Born type calculation
  where in gap states are found to appear easily in $s_{\pm}$ superconductors\cite{im1,im2}.

    In this paper we study the effect of non-magnetic impurities on two-band $s_{\pm}$-wave superconductors by analyzing
  the corresponding Ginzburg-Landau theory. The effect of impurities is included by generalizing the standard Bogoliubov
  de Gennes theory\cite{BdG} and diagrammatic perturbation techniques\cite{plee} which have been applied to study the
  effect of impurities on single-band superconductors\cite{BdG,whh,kotliar} to the case of two-band superconductors.

    We start with the Bogoliubov-de Gennes formulation of BCS theory\cite{BdG}. The system we consider is characterized by a
  BCS Hamiltonian, $H=H_0+V_{BCS}$, where
    \begin{subequations}
   \begin{equation}
   \label{h0}
     H_0=\sum_{i,j,\sigma}\int d^dr\psi^+_{i\sigma}(\vec{r})\left(\delta_{ij}\hat{H}_{0i}(\nabla)+U_{ij}(\vec{r})
     \right)\psi_{j\sigma}(\vec{r})
   \end{equation}
   where $i,j=1,2$ and $\sigma=\uparrow,\downarrow$ are the band and spin indices, respectively. $H_{0i}(\nabla)$ is the
   band Hamiltonian describing electronic wave-functions in band $i$. $U_{ij}(\vec{r})$ is a non-magnetic disordered
   potential which scatters electrons both within ($i=j$) and between ($i\neq j$) bands.
   $\psi_{j\sigma}(\psi^+_{j\sigma})$ are electron annihilation (creation) operators.

   \begin{equation}
   \label{V0}
   V_{BCS}=-\sum_{i,j,\sigma}V_{ij}\int d^dr
   \psi^+_{i\sigma}(\vec{r})\psi^+_{i\bar{\sigma}}(\vec{r})\psi_{j\bar{\sigma}}(\vec{r})\psi_{j\sigma}(\vec{r}),
   \end{equation}
   is the BCS interaction between electrons,  $\bar{\sigma}=-\sigma$. We note that $V_{ij}>0$ means
   attractive interaction in our notation.
   \end{subequations}
    Introducing the BCS decoupling,
    \[
   \psi^+_{i\sigma}\psi^+_{i\bar{\sigma}}\psi_{j\bar{\sigma}}\psi_{j\sigma}
   \sim\tilde{\Delta}^+_i\psi_{j\bar{\sigma}}\psi_{j\sigma}
   +\psi^+_{i\sigma}\psi^+_{i\bar{\sigma}}\tilde{\Delta}_j-\tilde{\Delta}^+_i\tilde{\Delta}_j,  \]
   where
   $\tilde{\Delta}_i=\langle\psi_{i\bar{\sigma}}\psi_{i\sigma}\rangle$
   we obtain the Bogoliubov-de Gennes equations for quasi-particle states $n$\cite{BdG},
   \begin{subequations}
   \label{bdg}
  \begin{eqnarray}
  \label{qeigen}
  \epsilon_nu_n^{(i)}(\vec{r}) & = & \sum_j\left(\delta_{ij}\hat{H}_{0i}(\nabla)+U_{ij}(\vec{r})
     \right)u_n^{(j)}+\Delta_i(\vec{r})v^{(i)}_n(\vec{r})
      \\ \nonumber
  \epsilon_nv_n^{(i)}(\vec{r}) & = & -\sum_j\left(\delta_{ij}\hat{H}^*_{0i}(\nabla)+U_{ij}(\vec{r})
     \right)v_n^{(j)}+\Delta_i^*(\vec{r})u^{(i)}_n(\vec{r})
  \end{eqnarray}
  where $\Delta_i(\vec{r})$'s are determined by the self-consistent equation,
  \begin{equation}
  \label{gap}
  \Delta_i(\vec{r})=-\sum_{j}V_{ij}\tilde{\Delta}_j(\vec{r})
  =\sum_{j,n}V_{ij}u_n^{(j)}(\vec{r})v_n^{(j)*}(\vec{r})(1-2f_n),
  \end{equation}
  \end{subequations}
  where $f_n=1/(e^{\beta(\epsilon_n-\mu)}+1)$ is the Fermi-Dirac distribution.

   We note that inter-band electron pairing is not included in our mean-field BCS decoupling. Physically different
 electronic bands describe electrons located at different parts of the Brillouin zone and an inter-band pairing
 implies finite-momentum Cooper pairs which is usually energetically not favorable. The mean-field decoupling we employed
 introduces only Josepshon coupling between superconducting order parameters in the two bands and the electronic
 wave-functions in the two bands are mixed only by the disorder-potential $U_{ij}$.

    The Ginzburg-Landau (GL) equation for the system can be derived by assuming that $\Delta_i(\vec{r})$ is small and
    expanding Eq.\ (\ref{bdg}) in powers of $\Delta_i(\vec{r})$ to third order. We furthermore assume that
    $\Delta_i(\vec{r})$ is slowly varying and perform a gradient expansion $\Delta_i(\vec{r}')\sim\Delta_i(\vec{r})
    +(\vec{r}'-\vec{r}).\nabla\Delta_i(\vec{r})+...$ to obtain\cite{BdG}
    \begin{eqnarray}
    \label{glb}
    \Delta_i(\vec{r}) & = & \sum_j\left(K_{ij}^{(0)}(\vec{r})\Delta_j(\vec{r})+{1\over2}K_{ij}^{(1)}(\vec{r})\nabla^2\Delta_j
    (\vec{r})\right)  \\ \nonumber
    & & +\sum_{jkl}L_{ijkl}^{(0)}(\vec{r})\Delta_j(\vec{r})\Delta_k^*(\vec{r})\Delta_l(\vec{r}),
    \end{eqnarray}
    where
    \begin{eqnarray}
    \label{ksa}
    K^{(n)}_{ij}(\vec{r}) & = & \int
    d^dr'(\vec{r}-\vec{r}')^{2n}K_{ij}(\vec{r},\vec{r}'), \\  \nonumber
    L_{ijkl}^{(0)}(\vec{r}) & = & \int d^dr_1\int d^dr_2\int d^dr_3L_{ijkl}(\vec{r},\vec{r}_1,\vec{r}_2,\vec{r}_3),
    \end{eqnarray}
    where
    \begin{subequations}
    \begin{equation}
    \label{k}
    K_{ij}(\vec{r},\vec{r}')=\sum_kV_{ik}\times{1\over\beta}\sum_{i\omega_n}g_{kj}(\vec{r},\vec{r}',i\omega_n)g_{kj}
    (\vec{r},\vec{r}',-i\omega_n),
    \end{equation}
    \begin{eqnarray}
    \label{L}
    & & L_{ijkl}(\vec{r}_1,\vec{r}_2,\vec{r}_3,\vec{r}_4)=-\sum_mV_{im}\times{1\over\beta}\sum_{i\omega_n}
    g_{mj}(\vec{r}_1,\vec{r}_2,i\omega_n)  \\ \nonumber
    & & g_{mk}(\vec{r}_1,\vec{r}_3,-i\omega_n)g_{lj}(\vec{r}_4,\vec{r}_2,i\omega_n)
    g_{lk}(\vec{r}_4,\vec{r}_3,-i\omega_n)
    \end{eqnarray}
    and
    \begin{equation}
    \label{g}
    g_{ij}(\vec{r},\vec{r}',i\omega_n)=\sum_n{\phi_n^{(i)}(\vec{r})\phi_n^{(j)*}(\vec{r}')\over
    i\omega_n-\epsilon_n}.
    \end{equation}
    \end{subequations}
    $\phi_n^{(i)}(\vec{r})$'s are eigenstates of $H_0$ given by
    $\epsilon_n\phi_n^{(i)}(\vec{r})=\sum_j
    \left(\delta_{ij}\hat{H}_{0i}(\nabla)+U_{ij}(\vec{r})\right)\phi_n^{(j)}(\vec{r})$.



     To study the effect of impurities we first consider the impurity-averaged GL equation where we replace $K^{(0)},K^{(1)}$ and
     $L^{(0)}$ by their averages over disorder potential $U_{ij}(\vec{r})$. Notice that we have assumed that
     $\langle K\Delta\rangle_{av}\sim\langle K\rangle_{av}\Delta$, etc. in this process where
     $\langle..\rangle_{av}$ denotes impurity average\cite{whh,kotliar}. The validity of this approximation will be examined
     later. With this approximation we obtain the usual impurity-averaged GL equation
     \begin{eqnarray}
    \label{GLav}
    \Delta_i(\vec{r}) & = & \sum_j\left(\bar{K}_{ij}^{(0)}\Delta_j(\vec{r})+{1\over2}\bar{K}_{ij}^{(1)}\nabla^2\Delta_j
    (\vec{r})\right)  \\ \nonumber
    & & +\sum_{jkl}\bar{L}_{ijkl}^{(0)}\Delta_j(\vec{r})\Delta_k^*(\vec{r})\Delta_l(\vec{r}),
    \end{eqnarray}
     where $\bar{K}^{(n)}=\langle K^{(n)}(\vec{r})\rangle_{av}$, etc.
     We shall consider the limit $E_f\tau>>1$ where $E_f\sim$ Fermi energy and $\tau\sim$ elastic scattering life time
     in our calculation and compute the impurity average to lowest order in impurity density $n_i$ (semi-classical limit)
     \cite{BdG,whh,kotliar}. In this limit electron motion becomes diffusive and modifies the long-distance
     behavior of $K_{ij}^{(n)}$ and $L_{ijkl}^{(0)}$.

     To compute $\langle K_{ij}(\vec{r},\vec{r}')\rangle_{av}$ we note that it can be written as\cite{kotliar}
   \begin{equation}
   \label{kav}
   \langle K_{ij}(\vec{r},\vec{r}')\rangle_{av}={1\over\beta}\sum_{k,i\omega_n}V_{ik}\int\int dEdE'
   {F_{kj}(E-E',\vec{r}-\vec{r}')\over(i\omega_n-E)(-i\omega_n-E')}
   \end{equation}
   where $F_{kj}(E-E',\vec{r}-\vec{r}')=\sum_{m,n}\langle\phi_n^{(k)}(\vec{r})\phi_n^{(j)*}(\vec{r}')
    \phi_m^{(k)}(\vec{r})\phi_m^{(j)*}(\vec{r}')\delta(E-\epsilon_n)\delta(E'-\epsilon_m)\rangle_{av}$
   is related to the density-density response function of the corresponding dirty metal\cite{plee},
   $\omega F_{ij}(\omega,\vec{q})=Im\chi_{ij}(\vec{q},\omega+i\delta)$,
   where $F_{ij}(\omega,\vec{q})$ is the Fourier transform of $F_{ij}(\omega,\vec{r})$ and
  $\chi_{ij}(\vec{q},\omega+i\delta)$ is the $(ij)$ component of the density-density response function of the dirty
  metal\cite{plee,kotliar}.

   The density-density response function can be evaluated to lowest order in impurity concentration by keeping the lowest
   order self-energy and particle-hole ladder diagrams\cite{plee}. To perform the impurity average we assume
   $\langle U_{ij}(\vec{r})\rangle_{av}=0$ and $\langle U_{ij}(\vec{r})U_{kl}(\vec{r}')\rangle_{av}\neq0$
   only if $i=k,j=l$ or $i=l, j=k$ with
    $\langle U_{ii}(\vec{r})U_{ii}(\vec{r}')\rangle_{av}=\delta^d(\vec{r}-\vec{r}')n_i|u_i|^2$ and
   $\langle U_{12(21)}(\vec{r})U_{12(21)}(\vec{r}')\rangle_{av}=\delta^d(\vec{r}-\vec{r}')n_i|u_t|^2$,
  i.e. the different type of scattering events are uncorrelated with each other. The corresponding
  averaged retarded (R) and advanced (A) electron Green's functions have the form\cite{plee}
   \[
   \langle
   g_{ij}^{R(A)}(\vec{k},\omega)\rangle_{av}={\delta_{ij}\over
   \omega-\xi_{i\vec{k}}+(-){i\over2\tau_i}}
    \]
   where $\tau_i^{-1}=\tau_{ii}^{-1}+\tau_{i\bar{i}}^{-1}$, $\tau_{ii}^{-1}=2n_i\pi|u_{i}|^2N_i(0)$ and
   $\tau_{i\bar{i}}^{-1}=2n_i\pi|u_t|^2N_{\bar{i}}(0)$, where $\bar{1}(\bar{2})=2(1)$ and $N_i(0)$ is the density of
   states for band $i$ electrons on the Fermi surface. $\tau_{ij}$ is the mean life time where an electron in a state
   in band $i$ is scattered to another state in band $j$.  Notice that the impurity-averaged Green's function has no
   off-diagonal $(i\neq j)$ term.

     The corresponding density-density response function is calculated to lowest order in $n_i$ by
   summing ladder diagrams in particle-hole channel (fig.1). We shall be interested at the low energy, long wave-length transport
   behaviors of the system. In this limit we need to keep only those processes where the particles and holes are
   coming from the same band in our calculation. This is because the two bands are located at different parts of
   the Brillouin zone, and the center of mass momentum of inter-band particle-hole excitations are usually large
   and do not contribute to small $\vec{q}$ processes.
  \begin{figure}
  \includegraphics[width=6cm, angle=0]{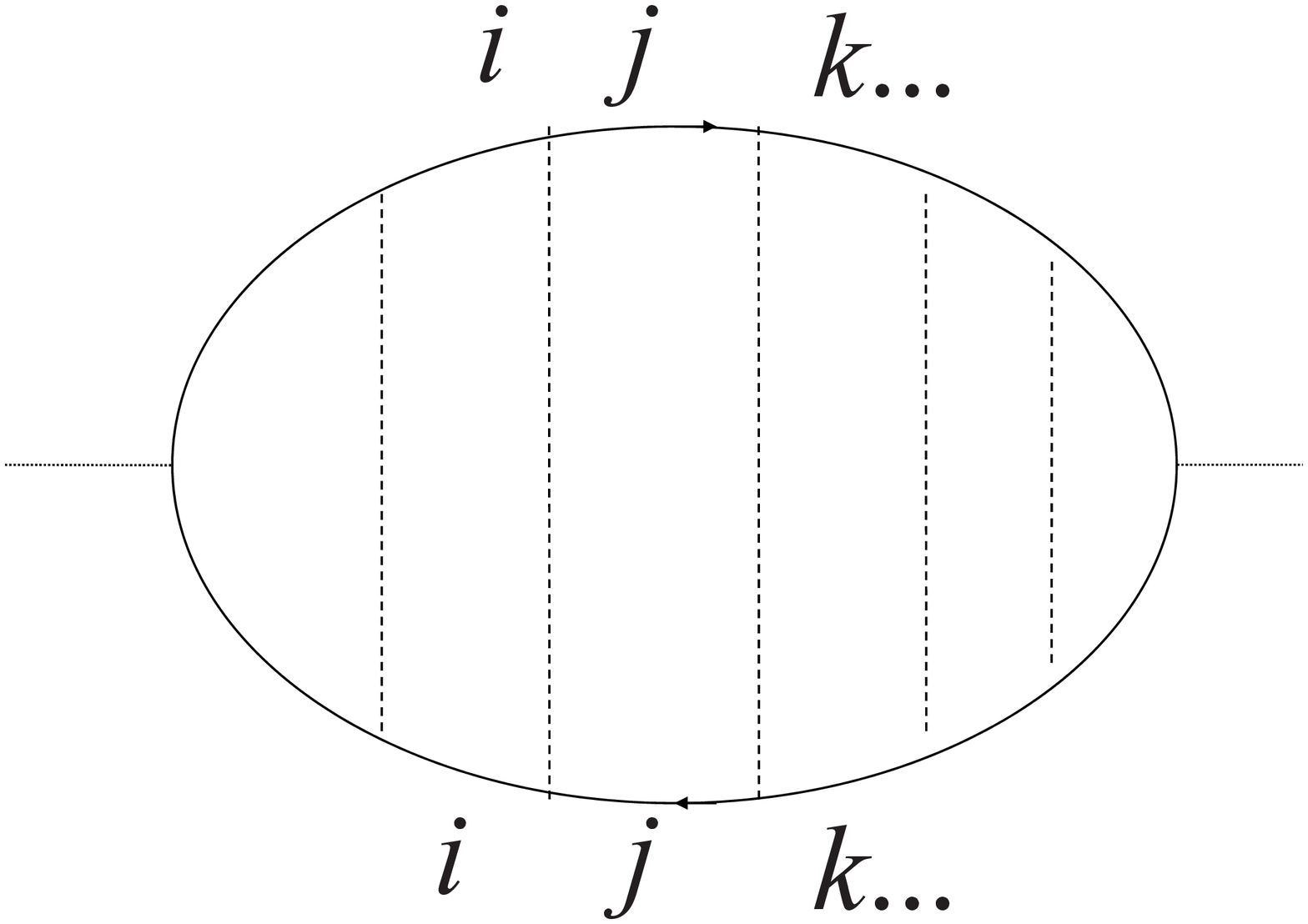}
  \caption{ladder diagrams in particle-hole channel, $i,j,k$ are band indices. We include only processes where particle and
  hole are coming from the same band.}
  \end{figure}

    Evaluating the diagrams, we obtain
    \begin{subequations}
    \label{chis}
    \begin{equation}
    \chi_{ii}(\vec{q},\omega)={(-i\omega+D_{\bar{i}}q^2+\tau_{\bar{i}i}^{-1})\rho_{ii}D_{i}q^2+
    \tau_{\bar{i}i}^{-1}\rho_{\bar{i}i}D_{\bar{i}}q^2\over
    (-i\omega+D_2q^2+\tau_{21}^{-1})(-i\omega+D_1q^2+\tau_{12}^{-1})-(\tau_{12}\tau_{21})^{-1}}
    \end{equation}
    and
    \begin{equation}
    \chi_{i\bar{i}}(\vec{q},\omega)={\tau_{\bar{i}i}^{-1}\left(\rho_{ii}D_iq^2+\rho_{\bar{i}i}D_{\bar{i}}q^2\right)\over
    (-i\omega+D_2q^2+\tau_{21}^{-1})(-i\omega+D_1q^2+\tau_{12}^{-1})-(\tau_{12}\tau_{21})^{-1}}
    \end{equation}
    \end{subequations}
    where $D_i=v_F^2\tau_i/d$ is the diffusion constant for band $i$ electrons and
    \begin{equation}
    \label{rij}
    \rho_{ij}=P_i(0)P_j(0)N_t(0), \quad~ P_{1(2)}(0)={N_{1(2)}(0)\over N_t(0)}.
    \end{equation}
    where $N_t(0)=N_1(0)+N_2(0)$ is the total density of states on the Fermi
    surface. The result is valid in the small $q,\omega$ limit $\omega<<\tau_i^{-1}$ and $D_iq^2<<\tau_i^{-1}$ for both
    $i=1,2$.


   $F_{ij}(q,\omega)$ can be evaluated using Eq.\ (\ref{chis}) and has very different behaviors at energy scales higher
   and lower than the inter-band scattering life-time $\tau_{i\bar{i}}^{-1}$. To simplify calculation we
   shall assume $\tau_{12}\sim\tau_{21}\sim\tau_t$ are of the same order of magnitude. In this case we obtain in the
   limit $\omega<<\tau_t^{-1}$ and $D_iq^2<<\tau_t^{-1}$,
  \begin{subequations}
  \label{fe}
  \begin{equation}
  \label{fl}
  F_{ij}(\vec{q},\omega)={\rho_{ij}D_{eff}q^2\over\omega^2+(D_{eff}q^2)^2},
  \end{equation}
  where $D_{eff}=P_1(0)D_1+P_2(0)D_2$ and
  \begin{equation}
  \label{fs}
  F_{ij}(\vec{q},\omega)=\delta_{ij}N_i(0){D_iq^2\over\omega^2+(D_iq^2)^2},
  \end{equation}
  \end{subequations}
  in the opposite limit $\omega>>\tau_t^{-1}$ and $D_iq^2>>\tau_t^{-1}$. Physically, electrons have scattered many times
  between the two band already in the limit $\omega,D_iq^2<<\tau_t^{-1}$ and the identity of bands
  is lost as far as electron dynamics is concerned. The only remaining information of ``bands" is that electrons have
  probability $P_i(0)$ of residing in band $i$. The identity of the two bands remain in the opposite limit
  $\omega,D_iq^2>>\tau_t^{-1}$ where electrons stay mainly in one-band. The two different limits expressed themselves in the
  GL equation where we find that in the limit $T_c<<\tau_t^{-1}$, electrons have to scatter
  between the two bands many times before forming a Cooper pair and the identity of intra-band Cooper pairs is lost,
  whereas intra-band Cooper pairs survived in the opposite limit $\tau_t^{-1}<<T_c$.
  We shall first consider the limit $T_c<<\tau_t^{-1}$ in the following.

  Putting together Eqs.\ (\ref{ksa}),\ (\ref{kav}) and\ (\ref{fe}) , we obtain
  \begin{eqnarray}
  \label{ks}
   \bar{K}^{(0)}_{ij} & \sim & \sum_kV_{ik}\rho_{kj}\ln{\omega_d\over T},  \\ \nonumber
  \bar{K}^{(1)}_{ij} & \sim & -\sum_k V_{ik}\times\rho_{kj}{D_{eff}\over  T_c}
  \end{eqnarray}
  where $\omega_d$ is the high energy cutoff for the attractive interaction in BCS theory. We have assumed $\tau_t^{-1}>>T_c$
  in deriving  $\bar{K}^{(1)}_{ij}$.

     $\bar{L}^{(0)}_{ijkl}$ can be computed similarly in perturbation theory. We obtain after some lengthy algebra
     \begin{equation}
     \label{k2}
     \bar{L}^{(0)}_{ijkl}\sim-{1\over
     T_c^2}\sum_mV_{im}P_m(0)P_j(0)P_k(0)P_l(0)N_t(0).
     \end{equation}
      The result can be understood most easily by noting that in the limit $T_c<<\tau_t^{-1}$ the dynamics of electron is
     described in an effective single-band picture with probability $P_i(0)$ of finding electrons in band $i$.

     Putting Eqs.\ (\ref{ksa}), \ (\ref{ks}) and\ (\ref{k2}) in Eq.\ (\ref{GLav}), multiple the resulting equation by $P_i(0)$
     and sum over $i$, we obtain an effective single band GL equation
     \begin{equation}
     \label{gleff}
     \Delta_{eff}(\vec{r})=a(T)\Delta_{eff}(\vec{r})-{b\over2}\nabla^2\Delta_{eff}(\vec{r})-c|\Delta_{eff}(\vec{r})|^2
     \Delta_{eff}(\vec{r})
     \end{equation}
     where $\Delta_{eff}(\vec{r})=\sum_iP_i(0)\Delta_i(\vec{r})$ and
     $a(T)=V_{av}N_t(0)\ln(\omega_d/T)$, $b\sim V_{av}N_t(0)D_{eff}/T_c$, $c\sim V_{av}N_t(0)/T_c^2$
     where $V_{av}=\sum_{ij}P_i(0)V_{ij}P_j(0)$ is the average interaction electrons see in forming the Cooper pairs.

     The individual band order parameters $\Delta_i(\vec{r})$ are related to
     $\Delta_{eff}(\vec{r})$ by
      \begin{equation}
      \label{eigen}
      \Delta_i(\vec{r})={1\over V_{av}}(\sum_kV_{ik}P_k(0))\Delta_{eff}(\vec{r}).
     \end{equation}
      and are `slaved' to $\Delta_{eff}$ in the sense that they are not independent dynamical variables in the system. The
     dirty two-band superconductor behaves as an effective dirty one-band superconductor in the regime $T_c<<\tau^{-1}_t$
     where measurement of superfluid properties cannot distinguish between whether the system is originally a single-band or
     a two-band superconductor.

    The effective single-band description has a number of interesting predictions. The (average) superconducting
    transition temperature is given by
    \begin{equation}
    \label{ei2}
    T_c=\omega_d\exp(-(V_{av}N_t(0))^{-1}).
    \end{equation}
    which is very different from clean two-band superconductors where $T_c$ is determined by
    \begin{equation}
    \label{ei1}
    T_c^{(0)}=\omega_d\exp(-\left({\bar{V}_{11}+\bar{V}_{22}\over2}+\sqrt{({\bar{V}_{11}-\bar{V}_{22}\over2})^2+|\bar{V}_t|^2}\right)^{-1}),
    \end{equation}
    where $\bar{V}_{ii}=V_{ii}N_i(0)$ and $\bar{V}_t=V_t\sqrt{N_1(0)N_2(0)}$ where $V_t=V_{12}=V_{21}$. Notice that
    $T_c^{(0)}$ is independent of $sgn(V_t)$.

     It is straightforward to show that $T_c\leq T_c^{(0)}$, i.e. $T_c$ is always lowered by disorder. However
     Eq.\ (\ref{ei2}) says that the precise value of $T_c$ is insensitive to the strength of disorder and depends only on the
     density of states of the two Fermi surfaces in the limit $\tau_t^{-1}<<T_c$! This surprising result is a direct
     consequence of ``Anderson Theorem"\cite{and} applied to the (effective) one-band superconductor.

     Contrary to the case of clean superconductors we also observe that $T_c$ depends now on the sign of $V_t$. In
     particular $T_c$ is enhanced by $V_t$ only if $V_t>0$, suggesting that disorder
     disfavor $s_{\pm}$ state. The relative sign between $\Delta_1$ and $\Delta_2$ depends on all the interactions
     now (Eq.\ (\ref{eigen})) and is not solely determined by $sgn(V_t)$!


       Next we consider the regime $\tau_t^{-1}\leq T_c$. This region is non-trivial as can be seen from the change in $T_c$ as
   a function of $\tau_t^{-1}$ determined by the GL theory. At $\tau_t^{-1}\rightarrow0$ $T_c$ is determined by
   Eq.\ (\ref{ei1}) for clean superconductors whereas $T_c$ is determined by Eq.\ (\ref{ei2}) at $\tau_t^{-1}>>T_c$.
   $T_c$ is {\em different but insensitive to disorder} at both regimes (Anderson Theorem)!
   Therefore Anderson Theorem must breaks down and $T_c$ becomes sensitive to disorder at the intermediate regime
   $0\leq\tau_t^{-1}\leq T_c$. The non-trivial effect of impurity scattering in this regime is shown in single-impurity
   calculations where it is found that in-gap bound states are induced easily by inter-band impurity scattering and the
   Josephson coupling between the bands is suppressed correspondingly in the $s_{\pm}$  state\cite{im1,im2,ng}. We note that
   the in-gap states are absent in the  $\tau_t^{-1}>>T_c$ limit where an effective single-band description becomes valid,
   consistent with findings on superconductors with sign-changing order-parameters\cite{preosti}.

        The rare (but strong) effects of inter-band impurity scattering suggests that the self-averaging
   approximation $\langle K\Delta\rangle_{av}\sim\langle K\rangle_{av}\Delta$ breaks down in the regime $\tau_t^{-1}\leq T_c$
   and $\Delta_i(\vec{r})$ becomes sensitive to the precise configuration of inter-band scattering  potentials. The
   sensitivity of $\Delta_i(\vec{r})$ to the impurity potential can also be seen
   directly from the (averaged) GL equation. It is easy to show that
 \[  \bar{K}^{(1)}_{ij}\rightarrow-V_{ij}N_j(0)D_i/T_c.
   \]
   and the GL equation does not take the form of an effective single-band GL equation in this regime. As a result its solutions
   are very sensitive to local variations in $K_{ij}^{(0)}(\vec{r})$.

     Therefore to describe the effects of order at this regime we should start with the un-averaged
    equation\ (\ref{glb}). It is more convenient is to replace the continuum GL equation by a random
    Josephson coupling lattice model with free energy
    \begin{eqnarray}
    \label{flat}
     F & = & \sum_i\left(\sum_{ml}a_{ml}(T;i)(\Delta^+_m(i)\Delta_l(i)+c.c.)+\sum_{m}b_{m}|\Delta_m(i)|^4\right)
     \\ \nonumber
     & & -\sum_{<ij>,l}t_{l}\left(\Delta^+_l(i)\Delta_l(j)+h.c.\right)
    \end{eqnarray}
     where $(i,j)$ and $(l,m)$ are lattice site and band indices, respectively. $<i,j>$ denotes nearest neighbor pair
     sites. The first term in\ (\ref{flat}) represents grains of two-band superconductors where the two bands are coupled only
     through Josephson coupling $a_{12(21)}$. The second term represents Josephson coupling between nearest neighbor grains.
     $a_{lm}(T;i)\rightarrow a_{lm}(T)$ with $a_{12(21)}>0$ for clean $s_{\pm}$ superconductors and $a_{lm}$ becomes randomized
     in the presence of disorder. It is easy to see from a three-site calculation that the phase of the order
     parameters are frustrated if $sgn(a_{12(21)}(T;i))$ becomes randomized\cite{nn}, indicating that a uniform
     superconducting state becomes unstable when inter-band impurity scattering is strong enough.

      Experimentally, we note that different superconducting gaps were observed at energy bands located at the $\Gamma-$ and
  $M-$ points of the pnictide superconductors in ARPES experiments\cite{ARPES2}, indicating that the materials are located
  in the weak inter-band scattering regime $\tau_t^{-1}\leq T_c$ where impurity-induced in-gap bound states are present,
  consistent with the existence of large density of in-gap states found in NMR\cite{NMR1} and lower critical
  field\cite{field} experiments. We propose here that a uniform superconducting state may become unstable at this regime. A
  detailed analysis of the superconducting behavior at this regime will be the subject of a separate paper.


\end{document}